\documentclass[preprintnumbers,amssymb, nofootinbib, nobibnotes,aps,prd]{revtex4}

\usepackage{amsmath,amssymb,amsfonts,amsthm}
\usepackage{soul, color}
\usepackage{graphicx}% Include figure files
\usepackage{dcolumn}% Align table columns on decimal point
\usepackage{bm}% bold math
\usepackage{t1enc}
\usepackage[english]{babel}
\usepackage[latin1]{inputenc}
\usepackage{color}
\usepackage{verbatim}
\usepackage{hyperref}
\usepackage{enumerate}

\theoremstyle{definition}

\usepackage{ulem}

\definecolor{bronze}{rgb}{0.8, 0.5, 0.2}

\renewcommand{\c}[1]{{\cal{#1}}}

\newcommand{\ra}{\rightarrow}

\newcommand{\be}{\begin{equation}}

\newcommand{\ee}{\end{equation}}

\emergencystretch=\maxdimen
\begin{document}

%\preprint{}

\title{Isoperimetric surfaces and area-angular momentum inequality in a rotating black hole in New Massive Gravity}

\author{Andr\'es Ace\~na}
 \email{acena.andres@conicet.gov.ar}
\affiliation{Facultad de Ciencias Exactas y Naturales, Universidad Nacional de Cuyo, CONICET, Mendoza, Argentina}
\affiliation{Observatorio Astron\'omico de Quito, Unidad de Gravitaci\'on y Cosmolog\'ia, Escuela Polit\'ecnica Nacional, Quito, Ecuador}
% \altaffiliation[Also at ]{Physics Department, XYZ University.}

\author{Ericson L\'opez}
 \email{ericsson.lopez@epn.edu.ec}
\affiliation{Observatorio Astron\'omico de Quito, Unidad de Gravitaci\'on y Cosmolog\'ia, Escuela Polit\'ecnica Nacional, Quito, Ecuador}

\author{Mario Llerena}
\affiliation{Observatorio Astron\'omico de Quito, Unidad de Gravitaci\'on y Cosmolog\'ia, Escuela Polit\'ecnica Nacional, Quito, Ecuador}

\date{\today}

\begin{abstract}
We study the existence and stability of isoperimetric surfaces in a family of rotating black holes in New Massive Gravity. We show that the stability of such surfaces is determined by the sign of the hair parameter. We use the isoperimetric surfaces to find a geometric inequality between the area and the angular momentum of the black hole, conjecturing geometric inequalities for more general black holes.
\end{abstract}

%\pacs{04.60.Kz,04.70.Bw,}
%\keywords{New Massive Gravity, rotating black hole, geometric inequalities}

\maketitle

%\tableofcontents

\section{Introduction}

Since its proposition in 2009 by Bergshoeff, Hohm and Townsend \cite{Berg}, New Massive Gravity (NMG) has received a great deal of attention, particularly due to its properties in the context of the AdS/CFT correspondence conjecture and because a variety of exact solutions have been found (see for example \cite{Berg2,Clement,Oliva}). The theory  describes gravity in a vacuum (2+1)-spacetime with a massive graviton. The action in this fourth-order derivative theory is given by
\begin{equation}
S=\dfrac{1}{16\pi G}\int d^3 x \,\sqrt{g} \left[R-2\lambda -\dfrac{1}{m^2}K\right],
\end{equation}
where $K=R_{\mu\nu}R^{\mu\nu}-\dfrac{3}{8}R^2$, while $m$ is a mass parameter. 
NMG admits solutions of constant curvature and possesses a unique maximally symmetric solution of constant curvature $\Lambda=2\lambda$ when $\lambda=m^2$. Static and stationary solutions have been found for this last case \cite{Oliva}. In the case of negative cosmological constant, the static solution found describes an asymptotically AdS black hole with a gravitational hair parameter. 

In this paper, we focus on the rotating solution also found in \cite{Oliva} that is also asymptotically AdS. It has a hair parameter and the rotational parameter satisfies $|a| < l$, where the parameter $l$ is related with the cosmological constant as $\Lambda=-1/l^2$. The extreme rotating case of this NMG black hole can be included after making a change in the hair parameter as suggested in \cite{Giribet}. The extreme case is obtained when $|a|=l$.  We are interested in the search of geometric inequalities as the one presented in \cite{Dain} for the Kerr black hole. We do this by finding the isoperimetric surfaces and analyzing their stability. This method has been applied to Reissner-Nordstr\"om in \cite{Acena2012}. Geometric inequalities are an important method to obtain physically relevant properties of metric theories, as they relate quantities of physical interest and tell us what type of phenomena is allowed within the theory.

The paper is organized as follows. In section \ref{sec:NMG} the family of rotating black holes in NMG is presented. Then the isoperimetric surfaces are found and the stability condition determined in section \ref{sec:isoperimetric}. The stable and unstable cases are determined in section  \ref{sec:Stability}. In section \ref{sec:conjecture}, we explore the geometric inequalities of area, mass and angular momentum for the NMG rotating black hole and conjecture inequalities for the general case. Finally, the conclusions are presented in section \ref{sec:conclusions}.

\section{The NMG rotating black hole}\label{sec:NMG}

As said, NMG is a theory that describes gravity in a vacuum (2+1)-spacetime with a massive graviton \cite{Berg}. An asymptotically AdS rotating black hole solution has been found in \cite{Oliva} and it contains a gravitational hair parameter $b'$. The rotational parameter $a$ is bounded by $-l<a<l$, where the parameter $l$ is related with the cosmological constant in the usual way, $\Lambda=-1/l^2$. The solution in the form presented in \cite{Oliva} does not include the extremely rotating case $|a|=l$, which we call from now on the extreme case. In order to include it, it is necessary to redefine the hair parameter as suggested in \cite{Giribet}. The new parameter is $b:= b'\xi^{-1}$, where $\xi$ is defined below. The rotating black hole solution that includes the extreme case is given by the metric  
\begin{equation}
\label{rotanteconb'}
ds^2=-{N}{F}dt^2+\dfrac{dr^2}{{F}}+r^2(d\phi+{N}^{\phi}dt)^2
\end{equation}
with
\begin{equation}
\label{1}
{N}=\left[1+\dfrac{{b}l^2}{4{\sigma}}(1-\xi)\right]^2,
\end{equation}
\begin{equation}
\label{2}
{N}^{\phi}=-\dfrac{a}{2r^2}(\mu-{b}{\sigma}),
\end{equation}
{\small
\begin{equation}
\label{3}
{F}=\dfrac{{\sigma}^2}{r^2}\left[\dfrac{{\sigma}^2}{l^2}+\dfrac{{b}}{2}(1+\xi){\sigma}+\dfrac{{b}^2 l^2}{16}(1-\xi)^2-\mu\xi\right],
\end{equation}}
\begin{equation}
\label{def_sigma}
{\sigma}=\left[r^2-\dfrac{\mu}{2}l^2(1-\xi)-\dfrac{{b}^2 l^4}{16}(1-\xi)^2\right]^{1/2},
\end{equation}
\begin{equation}
\label{5}
\xi^2=1-\dfrac{a^2}{l^2},
\end{equation}
where $\mu=4GM$, the angular momentum is given by $J=Ma$, $M$ is the mass measured with respect to the zero mass black hole and ${b}$ is the hair parameter. The rotational parameter $a$ satisfies $-l\leq a \leq l$ and the extreme case is obtained when $|a|=l$.

Before continuing, from \eqref{def_sigma} we see that $\sigma$ can be taken with either sign. If we allow it to be negative, then we notice that making the change $b\rightarrow -b$ and $\sigma\rightarrow-\sigma$ takes the metric functions to themselves, that is $N\rightarrow N$, $N^\phi\rightarrow N^\phi$, $F\rightarrow F$. Therefore, this change does not present new metrics, and we take only $\sigma$ as positive in this paper, which is also the right choice for the BTZ case (i.e. $b=0$).

These solutions possess one or more event horizons. If $b\leq 0$, the coordinate of the outermost horizon, $r_+$, is given by %\cite{AcenaEPN}
\begin{equation}
\label{r+}
r_+ = \frac{l}{\sqrt{8}}(1+\xi)^\frac{1}{2}\left[({b}^2l^2+4\mu)^\frac{1}{2} - {b}l\xi^\frac{1}{2}\right],
\end{equation}
and the parameters need to satisfy
\begin{equation}
\label{condMasa}
\mu \geq \mu_0 := -\dfrac{{b}^2l^2}{4}.
\end{equation}
The condition \eqref{condMasa} is presented in \cite{Giribet}. On the other hand, if $b>0$, the expression of the outermost horizon depends on the value of the mass, 
\begin{equation}
 r_+=\frac{l}{\sqrt{8}}(1+\xi)^\frac{1}{2}\left[({b}^2l^2+4\mu)^\frac{1}{2} - {b}l\xi^\frac{1}{2}\right]
 \label{r+1}\qquad\mbox{if}\qquad \mu\geq\mu_+,
\end{equation}
and
\begin{equation}
 r_+=\frac{l}{4}(1-\xi)^\frac{1}{2}\left[8\mu+{b}^2l^2(1-\xi)\right]^\frac{1}{2}\qquad\mbox{if}\qquad \mu_+\geq\mu\geq\mu_-,
 \label{r+2}
\end{equation}
where
\begin{equation}
\mu_+ := \frac{{b}^2l^2}{16}\frac{(1-\xi)^2}{\xi} \quad\mbox{and}\quad \mu_- := -\frac{{b}^2l^2}{8}(1-\xi).
\end{equation}
In general there is a curvature singularity, always hidden by the event horizon. Also, for $b\leq 0$, the extreme limit $|a|=l$ corresponds to a cylindrical end, in all similar to what happens for extreme Kerr. For details of this analysis the reader is refered to \cite{AcenaEPN}.

For these rotating solution we are interested in the search of geometric inequalities as the ones presented in \cite{Dain} for the Kerr black hole.

\section{Isoperimetric surfaces and stability condition}\label{sec:isoperimetric}

A hyper-surface in a manifold is isoperimetric if its area is an extreme with respect to nearby hyper-surfaces that enclose the same volume. This implies that the mean extrinsic curvature of an isoperimetric surface is constant \cite{Barbosa}. We are looking for isoperimetric surfaces in the slices of constant $t$ of the spacetime. We denote these slices as $\Sigma_{t_{0}}$ where $t_0$ is a constant. As the metric in  $\Sigma_{t_{0}}$ inherits the axial symmetry, then the hyper-surfaces of constant $r$ are necessarily isoperimetric. We denote these circles by $\Sigma_{t_0,r_0}$, with $r_0$ the radius of the isoperimetric surface in $\Sigma_{t_{0}}$. The calculation of the mean curvature \cite{Wald} of $\Sigma_{t_0,r_0}$ gives 
\begin{equation}
\chi=\dfrac{\sqrt{{F}(r_0)}}{r_0},
\end{equation}
which does not depend on the coordinate $\phi$ on $\Sigma_{t_0,r_0}$ and therefore confirms that $\Sigma_{t_0,r_0}$ is an isoperimetric surface in $\Sigma_{t_{0}}$.

An isoperimetric surface is called stable if its area is a minimum. For further discussion on isoperimetric surfaces in the context of General Relativity we refer to \cite{Dain:2011kb,dain12} and for the concept of stability to \cite{Barbosa}. We have the following condition on an isoperimetric surface $\Sigma$ to be stable \cite{Barbosa},
\begin{equation}
\label{generalcondition}
G(\alpha)\geq 0
\end{equation}
where
\begin{equation}\label{stabFun}
G(\alpha)=\int_{\Sigma}\left[-\alpha\Delta_{\Sigma}\alpha-\alpha^2(\chi_{AB}\chi^{AB}+R_{ab}n^an^b)\right]dA_{\Sigma}
\end{equation}
and $\alpha$ is any function on $\Sigma$ such that
\begin{equation}\label{condFunc}
\int_{\Sigma}\alpha\, dA_{\Sigma}=0,
\end{equation}
In \eqref{stabFun} $R_{ab}$ is the Ricci tensor in the Riemannian manifold, $n^a$ is the normal vector to the surface $\Sigma$, $\chi_{AB}$ is the extrinsic curvature of the surface $\Sigma$ and $dA_{\Sigma}$ is the volume element in $\Sigma$.

Evaluating $G(\alpha)$ for our case
\begin{eqnarray}
\nonumber G(\alpha) & = & \int_{\Sigma_{t_0,r_0}}\left[-\alpha\Delta_{\Sigma_{t_0,r_0}}\alpha-\alpha^2\left(\dfrac{{F(r_0)}}{r_0^{2}}-\dfrac{1}{2 \, r}\dfrac{\partial\,{F}}{\partial r}(r_0)\right)\right] dA_{\Sigma_{t_0,r_0}} \\
\label{change1}& = & \dfrac{1}{r_0}\int_0^{2\pi}\left[-\alpha\partial^2_{\phi}\alpha-\alpha^2\left({F}(r_0)-\dfrac{r_0}{2}\dfrac{\partial\,{F}}{\partial r}(r_0)\right)\right]\, d\phi.
\end{eqnarray}
We recall now that if $\Delta_{0}$ is the Laplace operator on the unit sphere, the eigenvalues $\lambda_k$ of the operator $-\Delta_{0}$ are given by $\lambda_k=k(k+n-1)$ where $k=0,1,...$ and $n$ is the dimension of the $n$-spheres \cite{Shubin}. 
Then in this case the first non-vanishing eigenvalue of the operator $-\partial^2_{\phi}$ is $\lambda_1=1$, which implies that
\begin{equation}
\int_0^{2\pi} -\alpha\partial^2_{\phi}\alpha \, d\phi \geq \int_0^{2\pi} \alpha^2 \, d\phi
\end{equation}
and therefore, from \eqref{change1},
\begin{equation}
G(\alpha)\geq\dfrac{1}{r_0}\left(1-{F(r_0)}+\dfrac{r_0}{2}\dfrac{\partial\,{F}}{\partial r}(r_0)\right)\int_0^{2\pi}\alpha^2\, d\phi.
\end{equation}
Considering the stability condition \eqref{generalcondition}, we have that $\Sigma_{t_0,r_0}$ is a stable isoperimetric surface if
\begin{equation}
\label{condestable1}
H(r_0):=1-{F(r_0)}+\dfrac{r_0}{2}\dfrac{\partial\,{F}}{\partial r}(r_0)\geq0.
\end{equation}
In the following we drop the subscript $0$ in $r_0$, and consider $r$ as a parameter that indicates which isoperimetric surface we are considering. What is left is to determine the stability of the isoperimetric surfaces, and this is performed in the following section, but before we write \eqref{condestable1} in two convenient forms. From \eqref{3} and \eqref{def_sigma} we have that $H$ can be written as
\begin{equation}
H=1-\left(1-\dfrac{A}{{\sigma}^2}\right){F}+\dfrac{{\sigma}}{l^2}({\sigma}+B)
\end{equation}
where
\begin{equation}
A=\dfrac{l^2}{16}(1-\xi)\left[{b}^2 l^2(1-\xi)+8\mu\right],\quad B=\dfrac{{b}l^2}{4}(1+\xi),\quad C=\dfrac{l^2}{4}\xi({b}^2l^2+4\mu),
\end{equation}
\begin{equation}
\label{F}
{\sigma}=\sqrt{r^2-A}, \qquad {F}=\dfrac{{\sigma}^2}{l^2 r^2}\left[({\sigma}+B)^2-C\right].
\end{equation}
$H$ can also be written as
\begin{equation}
H = \frac{1+\xi}{4r^2}h,
\end{equation}
with $h$ a third degree polynomial in $\sigma$,
\begin{eqnarray}
h & = & -{b}{\sigma}^3+\dfrac{4}{1+\xi}\left[1+\mu +\dfrac{{b}^2l^2}{16}(1-\xi)^2\right]{\sigma}^2 \nonumber \\
&& +\dfrac{3}{16}{b}l^2(1-\xi)\left[{b}^2l^2(1-\xi)+8\mu\right]{\sigma} \nonumber \\
&& +\dfrac{l^2}{4}\left(\dfrac{1-\xi}{1+\xi}\right)\left[{b}^2l^2(1-\xi)+8\mu\right]\left[1-\mu\xi+\dfrac{{b}^2l^2}{16}(1-\xi)^2\right],
\end{eqnarray}
and
\begin{equation}
H\geq 0 \iff h\geq 0.
\end{equation}

\section{Stability of the isoperimetric surfaces}\label{sec:Stability}

\subsection{Stability for the the BTZ black hole}\label{sec:BTZ}
If $b=0$ then the BTZ black hole with mass $\mu$ presented in \cite{Banados} is obtained. The stability condition \eqref{condestable1} is $H_{b=0}\geq0$ with
\begin{equation}
H_{b=0}=1+\mu-\dfrac{\mu^2 l^2}{2r^2}(1-\xi^2).
\end{equation}
For this black hole $\mu\geq 0$, so the stability condition is
\begin{equation}
r\geq r_c:=l\mu\sqrt{\frac{1-\xi^2}{2(1+\mu)}},
\end{equation}
where $r_c$ is the critical radial position for the isoperimetric surfaces. On the other hand, from \eqref{r+} the outer horizon is
\begin{equation}
r_{+}=l\sqrt{\frac{\mu}{2}(1+\xi)},
\end{equation}
therefore $r_+> r_c$ and all isoperimetric surfaces in BTZ are stable.

\subsection{Stability in the asymptotic region and at the horizon}\label{sec:stab-horizon}

As an intermediate step, we want to see if the isoperimetric surfaces are stable in the asymptotic region, $r\ra\infty$, and close to the horizon. For the asymptotic region, we see that $\sigma$ behaves as $r$, and therefore the leading order of the function $h$ is
\begin{equation}
 h \xrightarrow[r\ra\infty]{} -br^3,
\end{equation}
which gives that asymptotically
\begin{equation}
 h> 0 \mbox{ if } b< 0 \mbox{ and } h<0 \mbox{ if } b>0.
\end{equation}
So in the asymptotic region the isoperimetric surfaces are stable if $b<0$ and unstable if $b>0$.

At the horizon we have
\begin{equation}
 H_+ = 1-\left(1-\frac{A}{\sigma_+^2}\right)F(r_+)+\frac{\sigma_+}{l^2}(\sigma_++B),
\end{equation}
where a subscript $+$ indicates that the function is evaluated at $r_+$. Given that $F_+=0$, if $\sigma_+\neq0$ we have
\begin{equation}
 H_+ = 1 +\frac{\sigma_+}{l^2}(\sigma_++B).
\end{equation}
If $b<0$, then $\sigma_+>0$ and $\sigma_++B>0$, and therefore $H_+>0$. The same happens for $b>0$ and $\mu\geq\mu_+$. If $\mu_-<\mu<\mu_+$, then $\sigma_+=0$ and
\begin{equation}
h_+ = \dfrac{l^2}{4}\left(\dfrac{1-\xi}{1+\xi}\right)\left[{b}^2l^2(1-\xi)+8\mu\right]\left[1-\mu\xi+\dfrac{{b}^2l^2}{16}(1-\xi)^2\right] > 0.
\end{equation}
Given that all the involved functions are continuous, then there is always a neighborhood of $r_+$ where the isoperimetric surfaces are stable. It seems that the black hole stabilizes the isoperimetric surfaces in its neighborhood.

\subsection{Stability for $b<0$}

Due to the fact that for $b>0$ the isoperimetric surfaces in the asymptotic region are unstable, we focus on the case $b<0$, which also is the one that possess the cylindrical limit. To prove that all isoperimetric surfaces for $b<0$ are stable we perform the following steps. First we consider the function $h$ for the particular case of minimal mass, $\mu=\mu_0$. By taking its derivative with respect to $\sigma$ and analyzing its roots we show that it is an increasing function of $\sigma$, and this together with the stability near the horizon of the previous section shows that all isoperimetric surfaces for $\mu_0$ are stable. Then we consider the general case $\mu\geq\mu_0$, but this time we show that $h$ is an increasing function of $\mu$, and knowing from the previous step that for $\mu_0$ the function is positive it yields that all isoperimetric surfaces are stable for all allowed values of the parameters.

So we consider $b<0$ and $\mu=\mu_0$. Then
\begin{eqnarray}
h_{\mu=\mu_0} & = & -b\sigma^3 + \frac{4}{1+\xi}\left[1-\frac{b^2l^2}{16}(1+\xi)(3-\xi)\right]\sigma^2 \\
 && - \frac{3}{16}b^3l^4(1-\xi^2)\sigma - \frac{b^2l^4}{4}(1-\xi)\left[1+\frac{b^2l^2}{16}(1+\xi)^2\right].
\end{eqnarray}
We know that for $\sigma$ big enough this is an increasing function. The derivative with respect to $\sigma$ is
\begin{equation}
\partial_\sigma h_{\mu=\mu_0} = -3b\sigma^2 + \frac{8}{1+\xi}\left[1-\frac{b^2l^2}{16}(1+\xi)(3-\xi)\right]\sigma - \frac{3}{16}b^3l^4(1-\xi^2),
\end{equation}
and therefore the change from decreasing to increasing is at
\begin{equation}
 \sigma_c = \frac{-16+b^2l^2(1+\xi)(3-\xi)+16\sqrt{\Delta_1}}{12(-b)(1+\xi)},
\end{equation}
where
\begin{equation}
 \Delta_1 = 1-\frac{b^2l^2}{8}(1+\xi)(3-\xi)-\frac{b^4l^4}{128}\xi(1+\xi)^2(3-5\xi).
\end{equation}
It can be checked that
\begin{equation}
 \sigma_+(\mu_0) = -\frac{bl^2}{4}(1+\xi)>\sigma_c,
\end{equation}
and as $h_{\mu=\mu_0}(\sigma_+)>0$ then $h_{\mu=\mu_0}>0$ for $\sigma\geq\sigma_+$.

Now we consider $\mu\geq\mu_0$. The derivative of $h$ with respect to $\mu$ is
\begin{equation}
 \partial_\mu h = \frac{4}{1+\xi}\sigma^2 + \frac{3}{2}bl^2(1-\xi)\sigma + \frac{l^2}{8}\frac{1-\xi}{1+\xi}\left[16-32\mu\xi+b^2l^2(1-\xi)(1-3\xi)\right].
\end{equation}
The biggest root is
\begin{equation}
 \sigma_c = \frac{l}{16}\left[-3bl(1-\xi^2)+\sqrt{\Delta_2}\right],
\end{equation}
where
\begin{equation}
 \Delta_2 = (1-\xi)\left[-128+256\mu\xi+b^2l^2(1-\xi)(1+42\xi+9\xi^2)\right].
\end{equation}
We can check directly that $\sigma_+>\sigma_c$, and therefore $h$ is an increasing function of $\mu$ for $\sigma\geq\sigma_+$. This completes the proof that for $b<0$ all isoperimetric surfaces outside the horizon are stable.

\section{Geometric inequalities}\label{sec:conjecture}

We have shown that the surfaces $\Sigma_{t_0,r_0}$ are stable isoperimetric surfaces for $b\leq0$.  We now use these surfaces in the search of geometric inequalities. 

From the induced metric on $\Sigma_{t_0,r_0}$ we have that its area is simply
\begin{equation}
A=2\pi r_0.
\end{equation}
Then, the area of the horizon is $A_+=2\pi r_+$, and for $b\leq 0$ it takes the explicit form
\begin{equation}
A_+=\frac{\pi l}{\sqrt{2}}(1+\xi)^\frac{1}{2}\left[({b}^2l^2+4\mu)^\frac{1}{2} - {b}l\xi^\frac{1}{2}\right].
\label{AreaBH}
\end{equation}
Given that $r_+$ is the outermost horizon, then $r_0\geq r_+$, and therefore $A\geq A_+$. Also $A_+$ is an increasing function of $\xi$ and we have
\begin{equation}
 A_+ \geq A_e = \frac{\pi l}{\sqrt{2}}(b^2l^2+4\mu)^\frac{1}{2},
\label{ineqA-M}
\end{equation}
with equality only in the extreme case.

For the angular momentum we have $J = \dfrac{\mu a}{4G}$ with $|a|\leq l$. 
Now it is convenient to separate the analysis according to the sign of $\mu$. Let us start considering $\mu\geq 0$, then $\mu = \dfrac{4G}{l}|J_e|$, and
\begin{equation}
 |J_e|\geq |J|,
\end{equation}
where $J$ is the angular momentum of a spacetime where the other parameters are the same as in the extreme case. Putting all together we have
\begin{equation}
 A \geq A_+ \geq A_e = \frac{\pi l}{\sqrt{2}}\left(b^2l^2+\frac{16G}{l}|J_e|\right)^\frac{1}{2} \geq \frac{\pi l}{\sqrt{2}}\left(b^2l^2+\frac{16G}{l}|J|\right)^\frac{1}{2},
\end{equation}
therefore for any isoperimetric surfaces in a spacetime with $b\leq 0$ and $\mu\geq 0$
\begin{equation}
 A \geq \frac{\pi l}{\sqrt{2}}\left(b^2l^2+\frac{16G}{l}|J|\right)^\frac{1}{2}
 \label{inequality3}
\end{equation}
and the equality is only achieved in the extreme case. So, a black hole with a given area can not be rotating at any angular momentum because it has a maximal value depending on the hair parameter and the cosmological constant. For the BTZ black hole, from \eqref{inequality3}, we have that
\begin{equation}
A_{b=0} \geq \pi\sqrt{8Gl|J|}.
\end{equation}

On the other hand, for $\mu_0\leq \mu<0$,
\begin{equation}
 A \geq A_+ \geq A_e = \frac{\pi l}{\sqrt{2}}\left(b^2l^2+4\mu\right)^\frac{1}{2} = \frac{\pi l}{\sqrt{2}}\left(b^2l^2-4|\mu|\right)^\frac{1}{2}
\end{equation}
and given that $|\mu| = \dfrac{4G}{l}|J_e|$, so
\begin{equation}
 A_e = \frac{\pi l}{\sqrt{2}}\left(b^2l^2-\frac{16G}{l}|J_e|\right)^\frac{1}{2}.
\end{equation}
Here again, with the other parameters unchanged, $|J_e|\geq |J|$, so
\begin{equation}
 A_e \leq \frac{\pi l}{\sqrt{2}}\left(b^2l^2-\frac{16G}{l}|J|\right)^\frac{1}{2}.
\end{equation}
We see that in this case the area is not bounded by the angular momentum. In particular, for the extreme case with $\mu=\mu_0$ which has $A_e = 0$, it has a non-vanishing angular momentum given by $ |J_e| = \dfrac{b^2l^3}{16G}$.

In \cite{Giribet} and \cite{Giribet2010} it is proposed that the mass and angular momentum should be measured with respect to the extreme case with $\mu=\mu_0$. Accordingly, we redefine the mass and the angular momentum as follows
\begin{equation}
 \c{M} : = M - M_0 = \dfrac{1}{16 G}(b^2l^2+ 4 \mu),
\end{equation}
\begin{equation}
 \c{J} : = J - J_0 = Ma - M_0 a = \c{M}a,
\end{equation}
where $M_0 = \dfrac{\mu_0}{4G}$, and $J_0 = M_0 a$. It can be noticed that $\c{J}=J$ and $\c{M}=M$ in the BTZ case. These new parameters satisfy 
\begin{equation}
 \c{M}\geq 0,\qquad |\c{J}|\leq \c{M}l.
\end{equation}
Then the area of the extreme black hole can be expressed as
\begin{equation}
 A_e = \pi l \sqrt{8G\c{M}}.
\end{equation}
The angular momentum satisfies $|\c{J}_e| \geq |\c{J}|$ with $|\c{J}_e| = \c{M}l$, and fixing the other parameters, we finally have
\begin{equation}\label{areaIneq}
 A \geq \pi \sqrt{8Gl|\c{J}|}.
\end{equation}
It is indeed surprising that the definition of $\c{M}$ and $\c{J}$, which was motivated in \cite{Giribet} and \cite{Giribet2010} by black hole entropy considerations, is the right definition of mass and angular momentum regarding geometric inequalities.

To summarize and as conjecture for more general solutions of NMG, we have the following inequalities,
\begin{equation}
\c{M}\geq 0,
\end{equation}
\begin{equation}
|\c{J}|\leq \c{M}l,
\end{equation}
\begin{equation}
A \geq \pi \sqrt{8Gl|\c{J}|}.
\end{equation}

To complete the analysis we consider the case $b>0$. Here for $\mu<\mu_+$ we have that the horizon area is a decreasing function of $\xi$, which means that the minimum of the area is obtained for the minimum value of the angular momentum. Also, for $\mu<0$, the minimum angular momentum does not correspond to the static case (i.e. $\xi=1,\,J=0$) but to $\xi=1+\frac{8\mu}{b^2l^2}$, and the area of the horizon is zero (for details see \cite{AcenaEPN}). This situation can be remedied by redefining again the mass and angular momentum as
\begin{equation}
 \c{M} := M - M_-  = \frac{1}{4G}(\mu-\mu_-),
\end{equation}
\begin{equation}
 \c{J} := J - J_- = Ma - M_- a = \c{M}a,
\end{equation}
and they satisfy
\begin{equation}
 \c{M}\geq 0,\qquad -\c{M}l\leq \c{J}\leq \c{M}l.
\end{equation}

To simplify the notation we define the mass parameter
\begin{equation}
 \nu:=\frac{16G}{b^2l^2}\c{M}.
\end{equation}
The area of the horizon for the corresponding mass and angular momentum is
\begin{eqnarray}
 A_+ & = & \frac{\pi l^2b}{\sqrt{2}}(1-\xi)^\frac{1}{2}\nu^\frac{1}{2},\qquad 0<\nu<\nu_+,\\
 A_+ & = & \frac{\pi l^2b}{\sqrt{2}}(1+\xi)^\frac{1}{2}\left[\left(\nu+\frac{1+\xi}{2}\right)^\frac{1}{2}-\xi^\frac{1}{2}\right],\qquad \nu_+\leq\nu,
\end{eqnarray}
with
\begin{equation}
 \nu_+ = \frac{1-\xi^2}{4\xi}.
\end{equation}
We need to consider two ranges of the mass parameter. For $0<\nu<\nu_c$, the minimum of the area of the horizon with respect to the angular momentum once the mass and other parameters are fixed is obtained for $\xi=1$, that is, for the static case, where
\begin{equation}
 \nu_c \approx 5.64.
\end{equation}
For $\nu\geq\nu_c$, the minimum area is obtained for the angular momentum parameter satisfying
\begin{equation}\label{grad0}
 \nu = \frac{1+2\xi+2\xi^2+(1+2\xi)\sqrt{1+2\xi+2\xi^2}}{2\xi}.
\end{equation}
Therefore, the minimum of the area is never obtained in the extremely rotating case, and hence an inequality in the spirit of \eqref{areaIneq} can not be obtained. This can be expected from the solution not having a cylindrical end, which makes the extreme case a not typical extreme case, and from the isoperimetric surfaces in the constant$-t$ slices not being all stable.

\section{Conclusions}\label{sec:conclusions}

We have analyzed the existence and stability of isoperimetric surfaces in the $t=constant$ slices of the black hole solutions of NMG. We concluded that the determinant factor deciding the stability of the isoperimetric surfaces is the sign of the hair parameter $b$, being stable for $b\leq0$ and unstable for $b>0$. Also, for either sign of the hair parameter the isoperimetric surfaces are stable close to the horizon, which seems to indicate that the horizon performs a stabilizing function.

It needs to be pointed out that previous to the work \cite{AcenaEPN} particular attention to the case $b>0$ has not been paid. The conclusions of the present work support what was seen in \cite{AcenaEPN}, that the behavior of the solutions is radically different for $b>0$.

We have found geometric inequalities among the physical parameters of the solution. It is important to state that the particular selection of the physical parameters is crucial at this step, being surprising that a choice based on the analysis of thermodynamical properties (\cite{Giribet,Giribet2010}) is the one suited  to the geometric inequalities.

To continue the analysis, that is, to prove that the geometric inequalities conjectured hold for more general solutions of NMG, the constraint equations in the theory need to be analyzed. In that setting it would be particularly interesting to see how the hair parameter appears and how it is related to other stability properties.

\end{document}